\documentclass[aps,prl,amsmath,amsfonts,superscriptaddress,twocolumn]{revtex4}
\usepackage{epsfig,psfig,graphicx}

\newcommand{\beq}{\begin{equation}}
\newcommand{\enq}{\end{equation}}

\begin{document}

\title{Crossover temperature of Bose-Einstein condensation in an atomic Fermi gas}

\author{G.M. Falco}

\affiliation{Institute for Theoretical Physics, Utrecht
University, Leuvenlaan 4, 3584 CE Utrecht, The Netherlands}

\author{H.T.C. Stoof}

\affiliation{Institute for Theoretical Physics, Utrecht
University, Leuvenlaan 4, 3584 CE Utrecht, The Netherlands}

\begin{abstract}
We show that in an atomic Fermi gas near a Feshbach resonance the
crossover between a Bose-Einstein condensate of diatomic
molecules and a Bose-Einstein condensate of Cooper pairs occurs
at positive detuning, i.e., when the molecular energy level lies
in the two-atom continuum. We determine the crossover temperature
as a function of the applied magnetic field and find excellent
agreement with the experiment of Regal {\it et al.}
[Phys.~Rev.~Lett. {\bf 92}, 040403 (2004)] that has recently
observed this crossover temperature.
\end{abstract}


\maketitle

{\it Introduction.} --- An atomic Fermi gas near a Feshbach
resonance is a fundamentally new superfluid system. The reason is
that near a Feshbach resonance the gas does not only consist of
atoms but also of diatomic molecules. Moreover, the energy
difference between the molecular level and the threshold of the
two-atom continuum, known as the detuning $\delta$, can be
experimentally tuned by means of a magnetic field
\cite{stwalley1976,tiesinga1993}. In combination with the fact
that for fermionic atoms these molecules are very long lived
\cite{fermi0,fermi1,fermi2,fermi3}, such a gas thus offers the
exciting opportunity to study in detail the crossover between the
Bose-Einstein condensation (BEC) of diatomic molecules and the
Bose-Einstein condensation of atomic Cooper pairs, i.e., the
Bardeen-Cooper-Schrieffer (BCS) transition
\cite{stoof1996,timmermans2001,ohashi2002,milstein2002}. Indeed,
a Bose-Einstein condensate of molecules has recently been
observed \cite{jochim2003,greiner2003,zwierlein2003}. More
recently, a claim for Bose-Einstein condensation of atomic Cooper
pairs was made \cite{regal2004}. We will show, however, that the
date reported in Ref. \cite{regal2004} can be understood in terms
of a Bose-Einstein condensation of molecules.

At zero temperature the physics of the BEC-BCS crossover occurring
near a Feshbach resonance can be understood as follows. The
superfluid phase of the gas is always associated with a
Bose-Einstein condensate of pairs of atoms, but the wave function
of the pairs is given by
\begin{eqnarray}
\sqrt{Z(\delta)}\chi_{\rm m}({\bf x},{\bf x}') |{\rm closed}
\rangle + \sqrt{1-Z(\delta)}\chi_{\rm aa}({\bf x},{\bf x}';\delta)
                                               |{\rm open}\rangle~. \nonumber
\end{eqnarray}
At large negative detuning the energy of the molecule lies far
below the threshold of the two-atom continuum and we have
$Z(\delta) \simeq 1$. In that case we are dealing with a
Bose-Einstein condensate of diatomic molecules and the spatial
part of the pair wave function is equal to the (bare) molecular
wave function $\chi_{\rm m}({\bf x},{\bf x}')$. The spin part of
the pair wave function is then equal to $|{\rm closed} \rangle$,
i.e., the spin state of the closed channel of the relevant
Feshbach problem \cite{duine2003a}. At large positive detuning
the molecular energy level lies far above the threshold of the
two-atom continuum and can be (adiabatically) eliminated. We then
have that $Z(\delta) \simeq 0$ and the spatial part of the pair
wave function equals the usual BCS wave function for atomic
Cooper pairs $\chi_{\rm aa}({\bf x},{\bf x}';\delta)$. This
Cooper-pair wave function depends on the detuning, because the
effective attraction between the atoms depends on the detuning.
The spin state of the Cooper pairs is, however, always equal to
the spin state of the open channel of the Feshbach problem,
denoted here by $|{\rm open}\rangle$.

\begin{figure}[h]
\epsfig{figure=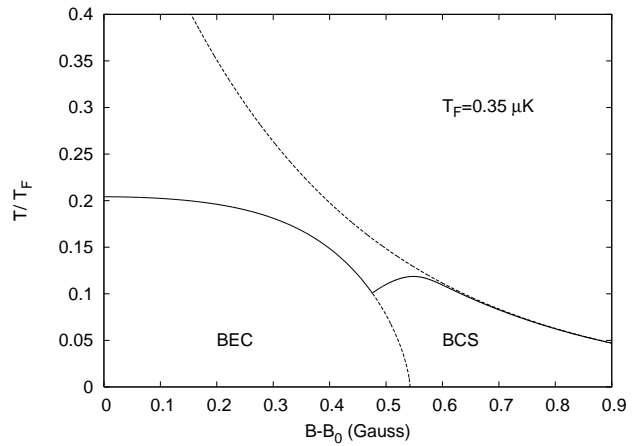,width=8.5cm} \caption{\rm Phase diagram
of atomic $^{40}$K as a function of magnetic field and
temperature for a Fermi temperature of the gas of $T_{\rm F} =
0.35~\mu$K. The solid line gives the critical temperature for
either a Bose-Einstein condensation of molecules or a
Bose-Einstein condensation of atomic Cooper pairs. The critical
temperature for the latter is calculated by simultaneously
solving the BCS gap equation and the equation of state of an ideal
mixture of atoms and molecules. In this equation of state, we use
the energy of the molecules given by Eq.~(\ref{energy}) below. For
comparison, the upper dashed curve is the analytical BCS result
$T/T_F = (8e^{\gamma-2}/\pi)e^{-\pi/2k_{\rm F}|a|}$, where $\gamma
\simeq 0.5772$ is Euler's constant \cite{stoof1996,melo1997}. The
lower dashed line is the crossover between the two Bose-Einstein
condensed phases, which is the main topic of this Letter.
\label{phasediagram}}
\end{figure}

With this physical picture in mind, the point where the crossover
takes place is thus determined by the detuning at which the
amplitude $Z(\delta)$ relatively abruptly crosses over from one
to zero. Based on two-body physics \cite{duine2003a} we would
expect the crossover to occur exactly on resonance, i.e., at zero
detuning. It is the main purpose of this Letter, however, to
point out that many-body physics changes this result and shifts
the crossover point to positive detuning. Quantitatively, the
crossover occurs, for an incoherent mixture of two hyperfine
states with equal density $n/2$, at the detuning where the
molecular energy level becomes equal to twice the Fermi energy
$\epsilon_{\rm F} = \hbar^2(3\pi^2n)^{2/3}/2m$. Using the theory
presented below we can also accurately determine the crossover
point at nonzero temperatures. The resulting (mean-field) phase
diagram for atomic $^{40}$K under the conditions of the
experiment of Regal {\it et al.} is shown in
Fig.~\ref{phasediagram} and summarizes the main conclusion of our
work.

Note that in the usual BEC-BCS crossover problem, studied in
condensed-matter physics in the context of the high-temperature
superconductors, $Z(\delta)$ is always identically zero and the
crossover is associated with a qualitatively different behavior
of the Cooper-pair wave function $\chi_{\rm aa}({\bf x},{\bf
x}';\delta)$ \cite{leggett1980,nozieres1985,melo1997}. This
emphasizes the fundamentally new nature of the superfluid state in
an atomic Fermi gas near a Feshbach resonance, which actually
shows a macroscopic coherence between atoms and molecules.

{\it Poor man's approach.} --- Before discussing the theory that
incorporates the resonant interactions between atoms, we first
consider the case of an ideal mixture of molecules and atoms to
establish the physical picture of the crossover most clearly. At
positive detuning a stable molecular state does not exist,
because the molecule can energetically decay into two free atoms.
In previous work \cite{duine2003a,duine2003b}, however, we have
shown that in first instance it is reasonably accurate to neglect
the finite lifetime of the molecule. Close to the Feshbach
resonance, the interaction with the atomic continuum shifts the
molecular energy level downward from the detuning $\delta$ to
$\epsilon_{\rm m} \simeq \hbar^2/ma^2$. Here $a(B)=a_{\rm
bg}[1-\Delta B/(B-B_0)]$ is the full atomic $s$-wave scattering
length of the Feshbach resonance, which is experimentally
characterized by its location at magnetic field $B_0$, its
magnetic field width $\Delta B$, and the background scattering
length $a_{\rm bg}$.

If the molecules are Bose-Einstein condensed, the chemical
potential of the atoms is equal to $\epsilon_{\rm m}/2 \simeq
\hbar^2/2ma^2$. If we take the atoms to be noninteracting, we can
thus at zero temperature immediately determine the density of
atoms in the Fermi sea below the chemical potential. Subtracting
this result from the total density $n$, and dividing by two, we
find that at zero temperature the density of molecules in the
Bose-Einstein condensate is equal to
\begin{equation}
n_{\rm mc} \simeq \frac{n}{2} \left[1 - \frac{1}{(k_{\rm F}|a|)^3}
\right]~, \label{molcond}
\end{equation}
where $k_{\rm F}=(3\pi^2n)^{1/3}$ is the Fermi momentum of the
gas. Note that this estimate is only valid for positive detuning,
where $a<0$. So at resonance the density of condensed molecules is
just n/2, i.e., half the total density of atoms in the gas.
Moreover, the density of condensed molecules vanishes at
$k_F|a|=1$. Physically, this situation occurs when the energy
level of the molecule is exactly equal to twice the Fermi energy
of the gas. This result is sensible, since if the energy level of
the molecule is higher, there will be no molecules at zero
temperature. The whole gas then consists of atoms. In the
experiment of Regal {\it et al.}, this condition gives a magnetic
field of 0.5 Gauss above the resonance, in excellent agreement
with the data shown in their Fig.~2.

To find the same criterion at nonzero temperatures is also
possible. We know that for temperatures below the Fermi
temperature the density of fermionic atoms is hardly influenced by
temperature. Not too close to resonance the molecular condensate
density is just depleted by thermal fluctuations, i.e., a thermal
cloud of molecules forms with increasing temperature. Calculating
the critical temperature for a density of ideal Bose molecules
given by Eq.~(\ref{molcond}) gives us the result
\begin{equation}
\frac{T}{T_{\rm F}} \simeq 2\pi \left\{ \frac{1}{6\pi^2\zeta(3/2)}
\left[1 - \frac{1}{(k_{\rm F}|a|)^3} \right] \right\}^{2/3}~,
\end{equation}
with $T_{\rm F}=\epsilon_{\rm F}/k_{\rm B}$ the Fermi temperature
and $\zeta(3/2) \simeq 2.612$. This result can be directly
compared with the data of Regal {\it et al.} presented in the x-y
plane of their Fig.~4. In view of the simplicity of the approach,
the agreement is remarkable.

{\it Molecular Bose-Einstein condensate.} --- To properly
incorporate the resonant interactions between the atoms, a more
involved treatment of the gas is necessary. Introducing creation
and annihilation operators for the molecules and atoms, the
grand-canonical hamiltonian of the gas becomes \cite{falco2004}
\begin{align}
\label{hamiltonian} H&=\int dx \psi^{\dagger}_{\rm m}(x)\left[
-\frac{\hbar^2\nabla^2}{4m}+ \epsilon_{\rm m}-2\mu \right]
\psi_{\rm m}(x) \\ &+ \sum_{\sigma=\uparrow,\downarrow} \int dx
\psi^{\dagger}_{\rm \sigma}(x)\left[
-\frac{\hbar^2\nabla^2}{2m}-\mu \right] \psi_{\rm \sigma}(x)
\nonumber
\\ &+\int dx g \left[ \psi^{\dagger}_{\rm m}(x)\psi_{\rm
\uparrow}(x)\psi_{\rm \downarrow}(x)+\psi^{\dagger}_{\rm
\downarrow}(x)\psi^{\dagger}_{\rm \uparrow}(x)\psi_{\rm m}(x)
\right]~, \nonumber
\end{align}
where the two hyperfine state of the atoms are denoted by
$|\uparrow\rangle$ and $|\downarrow\rangle$, the atom-molecule
coupling constant $g=\hbar\sqrt{4\pi a_{\rm bg} \Delta
B\Delta\mu_{\rm mag}/m}$, and the magnetic moment difference
$\Delta\mu_{\rm mag}$ between the hyperfine states $|{\rm
open}\rangle \equiv
(|\uparrow\downarrow\rangle-|\downarrow\uparrow\rangle)/\sqrt{2}$
and $|{\rm closed}\rangle$ gives the detuning $\delta =
\Delta\mu_{\rm mag}(B-B_0)$ \cite{singlet}. Finally, the
molecular energy is approximated by the energy where the
molecular density of states has a maximum. To find this maximum
we use that in the two-body T-matrix approximation the
frequency-dependent self-energy of the molecules is
$-i\eta\sqrt{\hbar\omega}$ \cite{duine2003a,falco2004}. In
general, this gives
\begin{align}
\epsilon_{\rm m}
=\frac{1}{3}\left(\delta-\frac{\eta^2}{2}+\sqrt{\frac{\eta^4}{4}
-\eta^2\delta+4\delta^2}\right)~, \label{energy}
\end{align}
where $\eta^2=g^4m^3/16\pi^2\hbar^6$ is the energy scale
associated with the width of the Feshbach resonance. This energy
scale is in fact of fundamental importance, because it shows that
at zero temperature the thermodynamic properties of a resonant
atomic Fermi are not solely determined by the Fermi energy. This
is particularly true for the experiment of Regal {\it et al.},
for which $\eta^2 \gg \epsilon_{\rm F}$, since for the Feshbach
resonance of interest $\eta^2 \simeq 7.7$ mK. Close to resonance,
where $\delta \ll \eta^2$, the molecular energy reduces to
$\delta^2/\eta^2$, which can be shown to be equivalent to
$\hbar^2/ma^2$ as expected. Note that the omission of the
background scattering length in the hamiltonian is justified
because the region of interest takes place relatively close to
resonance.

To find the crossover temperature for positive detuning, we
consider the gas to have a Bose-Einstein condensate of molecules,
and perform a quadratic expansion of the hamiltonian around the
nonzero expectation value $\langle \psi_{\rm m}({\bf x}) \rangle
\equiv \sqrt{n_{\rm mc}}$. This leads to the ideal gas expression
for the molecular density
\begin{equation}
n_{\rm m} = n_{\rm mc}+ \frac{1}{V} \sum_{\bf k}
\frac{1}{e^{\epsilon_{\bf k}/2k_{\rm B}T}-1}~,
\end{equation}
where $V$ is the volume of the gas and $\epsilon_{\bf
k}=\hbar^2{\bf k}^2/2m$. However, for atoms with momentum ${\bf
k}$, the resulting hamiltonian leads to a fluctuation matrix
\begin{align}
\left[
\begin{array}{cc}
\epsilon_{\bf k}-\epsilon_{\rm m}/2  & g \sqrt{n_{\rm mc}} \\ g
\sqrt{n_{\rm mc}} & -(\epsilon_{\bf k}-\epsilon_{\rm m}/2) \\
\end{array}
\right]~, \nonumber
\end{align}
which can easily be diagonalized by means of a Bogoliubov
transformation. Performing the calculation, we ultimately find
for the total atomic density
\begin{align}
n_{\rm a}&= \frac{2}{V} \sum_{\bf k} \Bigg(\frac{\epsilon_{\bf k}-
\epsilon_{\rm m}/2}{\hbar\omega_{\bf k}} \frac{1}{e^{\hbar
\omega_{\bf k}/k_{\rm B}T}+1} \nonumber \\
&+\frac{\hbar\omega_{\bf k}-\epsilon_{\bf k}+\epsilon_{\rm
m}/2}{2\hbar\omega_{\bf k}}\Bigg)~,
\end{align}
where the dispersion for the atoms obeys $\hbar\omega_{\bf
k}=\sqrt{(\epsilon_{\bf k}- \epsilon_{\rm m}/2)^2 + g^2n_{\rm
mc}}$.

\begin{figure}[h]
\epsfig{figure=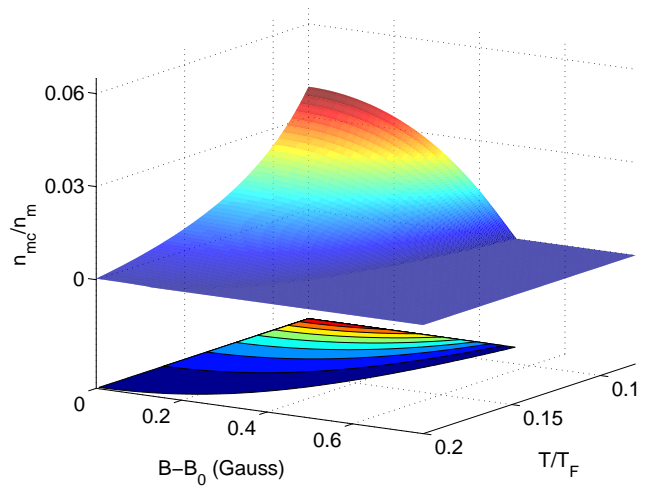,width=8.5cm} \caption{\rm Molecular
condensate fraction $n_{\rm mc}/n_{\rm m}$ in an atomic $^{40}$K
gas as a function of magnetic field and temperature for a Fermi
temperature of the gas of $T_{\rm F} = 0.35 \mu$K. This figure
should be compared with Fig.~4 of Ref.~\cite{regal2004}.
\label{condensate}}
\end{figure}

For fixed positive detuning and temperature, the equation for the
total atomic density $n=2n_{\rm m} + n_{\rm a}$ determines the
molecular condensate $n_{\rm mc}$. The result of these
calculations for the experiment of Regal {\it et al.} is shown in
Fig.~\ref{condensate}. This figure can be directly compared with
the data in their Fig.~4. Again the agreement is remarkable.
Having said that, it is important to realize that the experiment
is performed in an optical trap, whereas we have considered the
homogeneous situation. Generalizing our poor man's approach to
the trapped situation shows, however, that this does not affect
the position of the crossover line, because in that case the
homogeneous criterion is satisfied in the center of the trap. The
inhomogeneous analysis can be carried out in the local-density
(or Thomas-Fermi) approximation, but this is beyond the scope of
the present paper and is left for future work. Such an analysis
is certainly needed to obtain a full quantitative agreement
between theory and experiment.

The most important approximation that we have made in our
calculation of the crossover temperature is to neglect the finite
lifetime of the molecules. Including this finite lifetime is not
an easy task, because a consistent approach requires that the
self-energy of the molecules is calculated at least in the
many-body T-matrix approximation, just as we have done in our
recent work on the observation of molecular Kondo resonances in
an atomic Fermi gas near a Feshbach resonance \cite{falco2004}.
The physical reason for this complication is that the decay of
the molecules can be Pauli blocked by the presence of the atomic
Fermi sea. It is this Pauli blocking that is ultimately
responsible for the molecular Kondo resonances and it will,
therefore, also play an important role in a quantitative analysis
of the molecular lifetime effects. Qualitatively, we expect that
a broadening of the molecular energy level will not have a
substantial effect on the location of the crossover line. It
will, however, lead to an increase of the molecular condensate
fraction, because of the presence of molecular states with
energies below $\epsilon_{\rm m}$. We believe that these lifetime
effects may also have a bearing on the considerable narrowing of
the molecular thermal cloud that was also observed by Regal {\it
et al.} \cite{regal2004}. Work in this direction is presently
being completed and will be reported elsewhere.

We thank Randy Hulet for numerous stimulating discussions that
have lead us to the above. This work is supported by the Stichting
voor Fundamenteel Onderzoek der Materie (FOM) and the Nederlandse
Organisatie voor Wetenschaplijk Onderzoek (NWO).

\bibliographystyle{apsrev}

\begin{thebibliography}{104}
\bibitem{singlet} Since the label $\sigma$ of the hyperfine states
                  $|\sigma\rangle$ does not correspond to the
                  electron spin of the atom, the open channel of a
                  Feshbach resonance is actually never a singlet,
                  but usually contains mostly a triplet component.
                  As a result we almost always have that
                  $\Delta \mu_{\rm mag} \simeq 2 \mu_{\rm B}$, with
                  $\mu_{\rm B}$ the Bohr magneton.
\end{thebibliography}

\end{document}